\documentclass[twoside,slac_one]{revtex4}
\usepackage{graphicx}
\usepackage{fancyhdr}
\usepackage{amsmath} 
\usepackage{bm}
\usepackage{amsxtra}
\usepackage{amssymb}
\usepackage{amsthm}
\usepackage{latexsym}
\usepackage{lscape}

\begin{document}

\title{Combined CDF and D0 upper limits on MSSM Higgs boson production in proton-antiproton collisions at the Tevatron}

%

\author{L. Suter}
\affiliation{School of Physics and Astronomy, University of Manchester, Manchester, UK}

\begin{abstract}
We present combined results on the search for neutral supersymmetric Higgs bosons with data collected at the CDF and D0 experiments. Data were collected in proton-antiproton collisions at $\sqrt{s}$ = 1.96 ~TeV during Run II of the Tevatron. The searches considered cover the main production and decay mechanisms for Higgs bosons in tau lepton and bottom quark final states. The resulting combination is interpreted in the context of different scenarios within the Minimal Supersymmetric Standard Model.
\end{abstract}

\maketitle

\thispagestyle{fancy}


\section{Introduction}
Supersymmetry (SUSY) is an extension to the standard model (SM) that predicts an additional symmetry between bosons and fermions. This results in several advantages over the SM, such as the introduction of a dark matter candidate, a solution to the hierarchy problem and a potential for GUT scale unification.  The Minimal Supersymmetric Standard Model, MSSM,  ~\cite{MSSM} is the simplest version of a SUSY theory. It predicts, after symmetry breaking, five Higgs bosons of which three are neutral, these are denoted $h,H,A,H^+,H^-$.  At tree level the Higgs sector of the MSSM can be described by two parameters, which are chosen to be tan$\beta$, the ratio of the vacuum expectation values of the two Higgs doublets, and $M_A$, the mass of the pseudo-scalar Higgs boson. 
In the MSSM the coupling of the neutral Higgs to down-type quarks and leptons is enhanced by a factor of tan$\beta$ and the corresponding coupling to the up-type quarks and leptons is suppressed. This means that decay modes to bottom quarks and tau leptons are of specific interest, with a predicted decay rate of around 90\% to bottom quarks and 10\% to tau leptons. 
At large values of tan$\beta$ two of the three neutral Higgs bosons are approximately degenerate in mass which leads to an approximate doubling of the production cross section. 
Beyond tree level, radiative corrections bring in dependance on more than just $M_A$ and tan$\beta$. Therefore, limits are set for specific benchmark regimes in the $M_A$,  tan$\beta$ plane. 
For regions of the parameter space with low values of $M_A$ and large values of tan$\beta$ the Tevatron can set strong limits that complement the previous searches by the LEP experiments ~\cite{LEP}. 

\section{Combined CDF and D0 MSSM combination}

The combined results in the di-tau final state  from both D0 and CDF  are presented using up to 2.2 fb$^{-1}$ and 1.8 fb$^{-1}$ of integrated luminosity, respectively. The searches are described in ~\cite{old_comb1, old_comb2, old_comb3} and detailed information on the D0 and CDF detectors is available in ~\cite{D0, CDF}. 
The CDF analysis studied three final states, where either one tau decays leptonically to a electron or muon and one hadronically  or both  decay leptonically, $\tau_e\tau_{had}, \tau_{\mu} \tau_{had}$ and $\tau_{e} \tau_{\mu}$. At D0 a 1.0 fb$^{-1}$ analysis of these three final states was combined with an additional 1.2 fb$^{-1}$ analysis that studied just the $\tau_{\mu} \tau_{had}$ final state. More details on this combination are available in ~\cite{comb}.

\subsection{Analysis Summary} 

Electrons are identified through their characteristic energy deposits in the calorimeter and are required to be isolated and have a reconstructed track matching this energy deposit deposition in the calorimeter. For reconstructed muon objects, hits in the muon chambers are matched to tracks in the central tracking chambers. These are then required to be isolated in both the calorimeter and the central tracking detectors. 
D0 and CDF have different tools to identify hadronic decays of taus. At CDF narrow isolated clusters in the calorimeter with either one or three tracks are identified and a variable cone algorithm is used to reconstruct these with strict isolation requirements in order to suppress quark and gluon jets.
Hadronic tau decays at D0 are first split into three types, based on the energy clusters in the calorimeters and the number of tracks. Type 1 and 2 are 1-prong decays with energy deposited in the hadronic calorimeter (type 1) or in both the electromagnetic and hadronic calorimeters (type 2). Type 3 taus are 3-prong decays with an invariant mass below 1.7 GeV and energy deposits in the calorimeters. For each of the different types a Neural Network, NN,  is trained, using $Z \rightarrow \tau \tau$ decays as signal and multijet events predicted from data as background, to separate hadronic tau decays from jets.  An additional NN  trained on electron Monte Carlo events is used to remove backgrounds from electrons faking type 2 taus. 

 Monte Carlo simulations are used to determine the acceptance of the signal. The PYTHIA ~\cite{pythia} event generator is used with CTEQ5L parton sets ~\cite{parton} at CDF and CTEQ6L at D0. TAUOLA ~\cite{tauola} is used to simulate the final state tau lepton decays.  For CDF both $gg$ and $b\bar{b}$ production modes are considered, D0 just considers  $gg$ production. The acceptances are very similar for both production modes. The response of the detector is modeled by GEANT ~\cite{geant}, the SM backgrounds are generated by PYTHIA, except $t\bar{t}$ which is generated by COMPHEP with PYTHIA ~\cite{comphep} and W/Z samples with additional jets in the sample which is generated by ALPGEN ~\cite{alpgen} with PYTHIA used as normalization.  Diboson and $t\bar{t}$ are normalized to next-to-leading order ~\cite{NLO1, NLO2} and $Z/ \gamma^*$ samples are normalized to next-to-next-to-leading order ~\cite{NNLO}. 
D0 uses a inclusive electron and muon trigger and CDF a lepton plus track trigger. After reconstruction, candidate events must have two isolated leptons of opposite charge. The main backgrounds are $Z \rightarrow \tau \tau, \mu\mu, ee$, multijet, $W \rightarrow e\nu, \mu\nu, \tau\nu$, diboson and $t\bar{t}$.
For the $\tau_{e}\tau_{had}$ and $\tau_{\mu}\tau_{had}$ channels the following cuts are applied at CDF;  the muon or electron is required to be isolated with a  $p_T$ cut of $>$ 10 GeV, the tau is required to have $p_T >$ 15 GeV for a one-prong tau and $>$ 20 GeV for three-prong taus.
At D0 the muon or electron is required to be isolated with a  $p_T$ cut of $>$ 15 GeV , the tau is required to have $p_T >$ 16.5 GeV for a one-prong tau and $>$ 22 GeV for a three-prong tau. 
  Additional cuts are placed on the relative directions of the tau, the ${E\kern -0.55em/_{T}}$ and the transverse mass. At CDF a cut is also placed on the scalar sum of the transverse momenta, defined as $H_T= | p_{T}^{e/\mu} | + |  p_{T}^{\tau_{had}} | + | {E\kern -0.55em/_{T}} | > 55$ GeV. 
For the $\tau_{e}\tau_{\mu}$ analysis, events with a central muon and electron are selected, with min$(E_{T}^{e}, p_{T}^{\mu}) > 6$ GeV, max$(E_{T}^{e}, p_{T}^{\mu}) > 10$ GeV  and $|E_{T}^{e}| + |p_{T}^{\mu}| > 30$ GeV. At D0 it is required that $p_{T}^{\mu} > 10 $GeV, $p_{T}^{e} > 12$ GeV, invariant mass of the $e\mu$ pair $>$ 20 GeV and $|E_{T}^{e}| + |p_{T}^{\mu}| + | {E\kern -0.55em/_{T}}| > 65$ GeV. 
The expected yields for signal and background events can be seen in Table I along with the observed data events.

\begin{table*}[t]
\begin{center}
\caption{Expected numbers of background and signal events for $\tau\tau$ final states.}
\begin{tabular}{|l|ccc||l|ccc|}
\hline \textbf{CDF} L = 1.8 fb$^{-1}$ &\textbf{$\tau_e\tau_{\mu}$} & 
\textbf{$\tau_e\tau_{had}$}&
\textbf{$\tau_{\mu}\tau_{had}$}&
\textbf{D0} L $ < $ 2.2 fb$^{-1}$ &\textbf{ $\tau_e\tau_{\mu}$} &
\textbf{$\tau_e\tau_{had}$} &
\textbf{$\tau_{mu}\tau_{had}$} \\
\hline 
$Z\rightarrow \tau\tau$ & 605 $\pm$ 51 & 1378 $\pm$ 117 & 1353 $\pm$ 116 &
 & 212 $\pm$ 20 & 581 $\pm$ 5  & 2153 $\pm$ 156  \\
$Z\rightarrow ee/Z\rightarrow \mu\mu$ & 19.4 $\pm$ 5.7 & 70 $\pm$ 10 &  107 $\pm$ 13&
 & 10.4 $\pm$ 1.3 & 31 $\pm$ 2  & 66 $\pm$ 8  \\
diboson$ + t\bar{t}$ & 20.5 $\pm$ 7.0  & 8.2 $\pm$ 4.2 & 6.6 $\pm$ 3.7 & 
 & 6.1 $\pm$ 0.6 & 3.1 $\pm$ 0.3  & 16 $\pm$ 3  \\
multijet$ + W \rightarrow l\nu$ & 57.1 $\pm$ 13.5 & 467 $\pm$ 73 & 285 $\pm$ 46  &
 & 37.9 $\pm$ 7.7 & 374 $\pm$ 48  & 216 $\pm$ 41 \\
\hline
Total Background  & 702 $\pm$ 55 & 1922 $\pm$ 141 & 1752 $\pm$ 129  &
 & 266 $\pm$ 22 & 989 $\pm$ 82  & 2451 $\pm$ 162 \\
\hline
Data & 726 & 1979 & 1666  &
 & 274 & 1034  & 2340  \\
\hline
\end{tabular}
\label{example_table_2col}
\end{center}
\end{table*}



\subsection{Combination}

Two different methods are used to set the limits to be confident that the final results do not depend on the details of the statistical formulation.  These are the Bayesian and Modified Frequentist approaches and rely not just on the integrated yields of the analyses but on the shapes of the final discriminants. Systematic uncertainties appear as both uncertainties on the number of expected signal and background and distributions of the final discriminates. For both D0 and CDF the visible mass  defined as the invariant mass of the visible tau decay products and ${E\kern -0.55em/_{T}}$ was used as the final discriminant.

\subsection{Systematic Uncertainties} 

The systematic uncertainties that are correlated between the two experiments include the uncertainties on the integrated luminosity, on the rate for $t\bar{t}$ production, signal and diboson production.
As both CDF and D0 extract the multijet background from the data using two different techniques, the uncertainties on the multijet prediction is uncorrelated between the two experiments. Similarly the methods of calibrating the fake lepton rates, b-tag efficiencies and mistag rates are performed separately by each experiment and are uncorrelated.

\subsection{Combined Results}

The two methods used to set the limits show good agreement with each other with deviations at less than the 10\% level. These results can be seen in Fig. 1 using the $CL_s$ method over the range $ 90 < M_A < 200$ GeV. The observed limits are in good agreement with the expectation with no significant excess observed. 

\begin{figure}[ht]
\centering
\includegraphics[width=80mm]{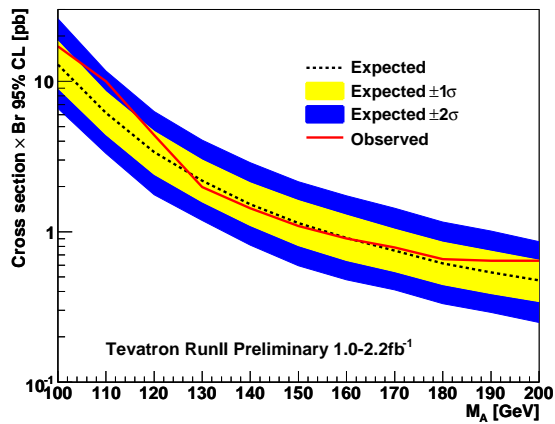}
\caption{95\% confidence limits on the cross section $\times$ branching ratio of a neutral MSSM Higgs boson. The solid and dashed lines show the observed and expected limits respectively. The yellow and blue bands show the 1$\sigma$ and 2$\sigma$ bands.  } \label{figure_one}
\end{figure}

These limits can be interpreted within the MSSM in specific benchmark scenarios. Four specific scenarios are defined by setting the values of $M_{SUSY}$, the mass scale of the quarks, $\mu$, the Higgs sector bilinear coupling, $M_2$, the gaugino mass term, $A_t$, the trilinear coupling of the stop sector, $A_b$,  the trilinear coupling of the sbottom sector and $m_g$, the gluino mass term. The maximal mixing scenario $m_h^{max}$ is defined by setting 

\begin{equation} \begin{split}
& M_{SUSY} = 1 \mathrm{~TeV},\mathrm{~}\mu = 200  \mathrm{~GeV},\mathrm{~} M_2 = 200  \mathrm{~GeV}\\
&  X_t = 2M_{SUSY},\mathrm{~} A_b = A_t,\mathrm{~} m_g = 0.8M_{SUSY}.  \\
\label{eq-sp}
\end{split}
\end{equation}
 
\noindent The no-mixing scenario is defined by requiring $ M_{SUSY} = 2$  TeV and $X_{t} = 0$, with the rest of the parameters unchanged from the $m_h^{max}$ scenario. The two additional scenarios are created be changing the sign of the $\mu$ parameter. The 95\% confidence limits set in these scenarios can be seen in Figure 2 for the maximal and no mixing scenarios and a positive $\mu$ parameter. FEYNHIGGS ~\cite{feynhiggs} was used to calculate the signal cross section and the branching fraction with no theoretical uncertainties considered. 
For large values of tan$\beta$ the width of the Higgs boson can be compared to the experimental resolution. This width dependance is not taken into account when calculating the limits but in the region where the limits are set it is not expected to be a large effect ~\cite{width}.

\begin{figure*}[ht]
\centering
\includegraphics[width=80mm]{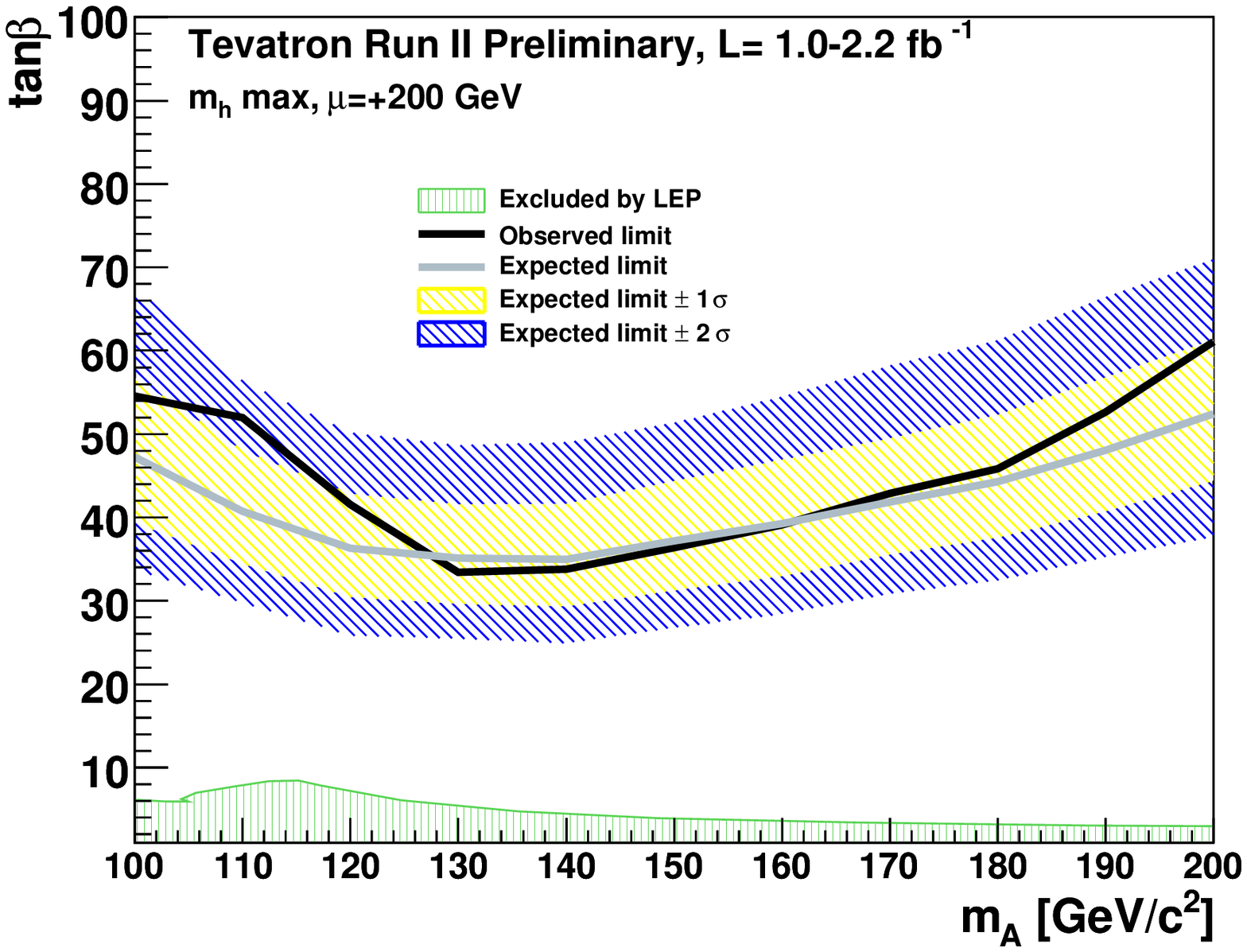}
\includegraphics[width=80mm]{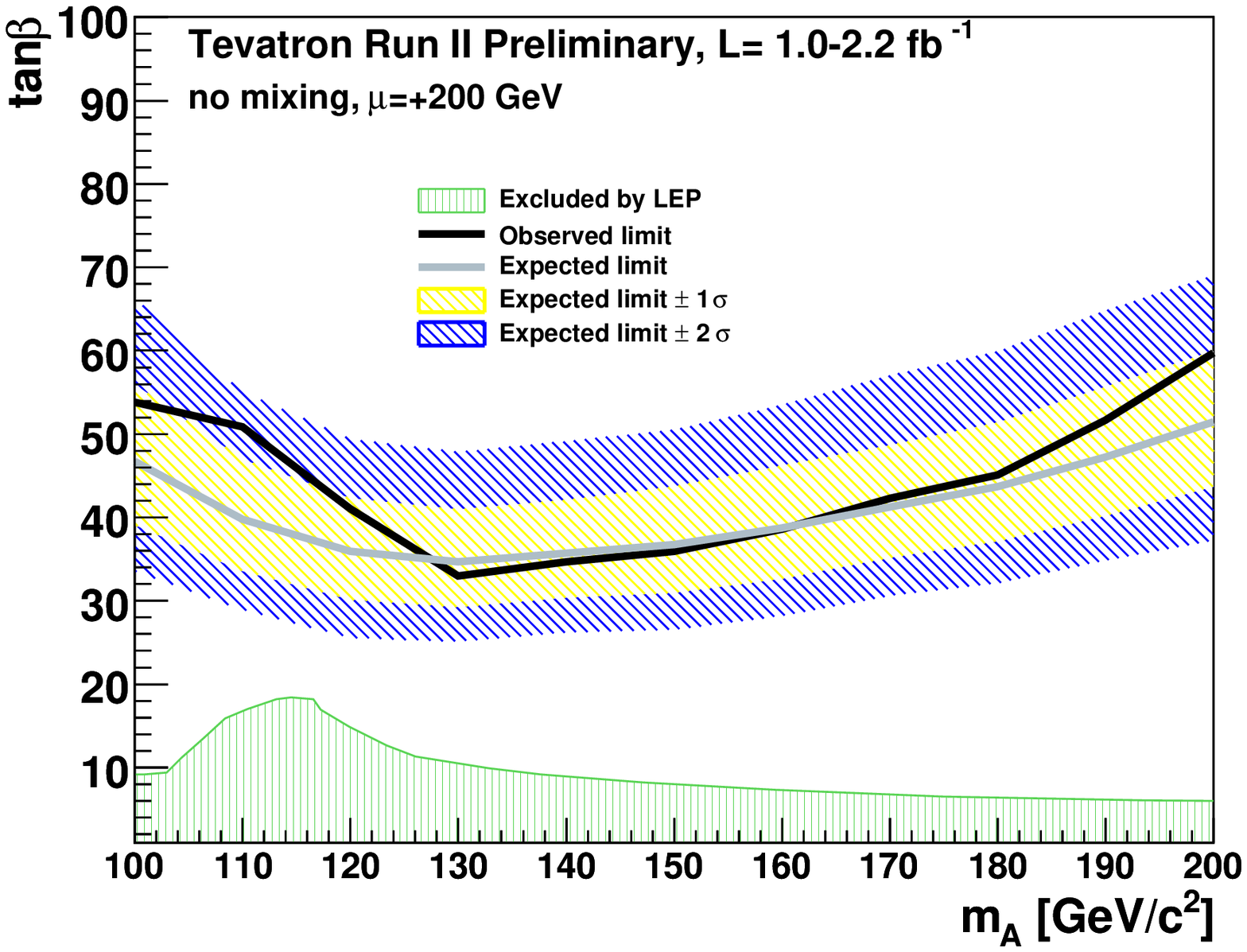}
\caption{95\% confidence limits in the $M_A$ tan$\beta$ plane, for the two benchmark scenarios and  $\mu >$ 0. The black line denotes the observed limit, the grey line the expected limit and the hatched yellow and blue regions the 1$\sigma$ and 2$\sigma$ bands. The green region is the region excluded by LEP. } \label{figure_two}
\end{figure*}

\section{D0 MSSM combination}

The combined results from D0 on the search of a neutral Higgs boson produced in association with a $b$ quark, in the $bh \rightarrow b\tau\tau$ and $bh \rightarrow bb\bar{b}$ channels, using an integrated luminosity of 7.3 fb$^{-1}$ and 5.2 fb$^{-1}$, respectively, is presented.  Limits are set at the 95\% confidence level on the production cross section of a neutral Higgs boson produced with one or more $b$ quarks, $\sigma(gb \rightarrow hb(b))$, for Higgs boson masses $90 < M_A < 300$ GeV, where it is assumed that $BR(h\rightarrow bb) + BR(h \rightarrow \tau\tau)$ = 1 and there is a narrow Higgs width.  This combination is described in detail in ~\cite{d0comb}. 

\subsection{Analysis Summary} 

Detailed descriptions of the two combined analyses $bh \rightarrow b\tau\tau$ and $bh \rightarrow bb\bar{b}$ are available in ~\cite{bana1, bana2}. 
The muon and hadronic tau identification is performed as described in Section 2.1. For an object to be selected as a $b$-jet candidate it must pass as set of quality criteria. Within a cone of $\Delta R = \sqrt{ (\Delta \eta)^2 + (\Delta \phi)^2} < 0.5$ around the jet axis it must have two or more tracks. A Neural Network algorithm trained on lifetime information from the track input parameters and secondary vertices is then used to further select events ~\cite{btag}. For more information on jet reconstruction and energy scale determination, see ~\cite{jets}. 

To determine the signal acceptance, the $gh \rightarrow hb(b)$ process is simulated using PYTHIA in the 5-flavor scheme at leading order. NLO corrections are calculated using MCFM ~\cite{mcfm} and are applied as weights on the transverse momentum, $p_T$, and the $\eta$  distributions. 
TAUOLA  is used to simulate the final state tau lepton decays and GEANT to model the detector response. Weights are applied to correct for the differences between the actual detector response and the simulation. 
All backgrounds are simulated by ALPGEN, with PYTHIA used to model the hadronization and showering, apart from the diboson processes which are simulated by  PYTHIA only. The diboson and $t\bar{t}$ samples are normalized to next-to-leading order, whereas $Z/ \gamma^*$ samples are normalized to next-to-next-to-leading order.  

\subsubsection{ $bh \rightarrow b\tau\tau$}

For the $bh \rightarrow b\tau\tau$ analysis, where one tau is required to decay hadronically and one to a muon, an inclusive trigger approach is used which is a combination of the signal muon, electron, jet, tau, muon + jet and muon + tau triggers. One isolated muon is required with a matching central track, which must pass the cuts: $p_T > 15 $ GeV, $|\eta_{det}| < 1.6$ . Events with greater than one muon are rejected to suppress the $Z \rightarrow \mu\mu$ background.  
The hadronic tau candidates must be isolated with the cuts: $p_T > 15 $ GeV, $|\eta_{\tau}| < 2.5$. A cut is also applied on the tau Neural Network output, NN$_{\tau}$, which selects hadronic tau candidates with a efficiency of 65\% while rejecting jets with a efficiency of around 99\%. 
Events must also have a $b$-jet which passes the following criteria; isolated from the tau and muon, $p_T > 15$ GeV, $|\eta| < 2.5, |\eta_{det}| < 2.5$. In addition a cut is made on the $b$ tagging Neural Network, NN$_b$, which selects $b$-jet candidates with an efficiency of around 65\% and a fake rate of 5\%.  
The main backgrounds are multijet, $t\bar{t}$ and $Z \rightarrow \tau\tau$ produced with heavy flavor jets, with small contributions from $W$+jets, $Z\rightarrow \tau\tau$ plus light jets, $Z \rightarrow ll$, single top production and diboson production. For $W$+jets the Monte Carlo is normalized to data using a control sample.  Multijet background is estimated from data using inverted cuts to select a multijet-rich sample. 
To further reject the main backgrounds, multijet and $t\bar{t}$. A discriminant is trained on the multijet background, $D_{MJ}$, and a NN on the $t\bar{t}$ background, $D_{t\bar{t}}$. The final discriminant that is used for the limit setting is a likelihood  discriminant, $D_f$, which is constructed from $D_{MJ}$ and $D_{t\bar{t}}$ as well as the $b$-tagging NN, and a variable relating to the decay kinematics, $M_{hat}$. The distribution for $D_f$ is shown in Figure 3 for all tau types and a 110 GeV Higgs signal (for the statistical analysis the tau types are treated separately). 

\begin{figure}[ht]
\centering
\includegraphics[width=80mm]{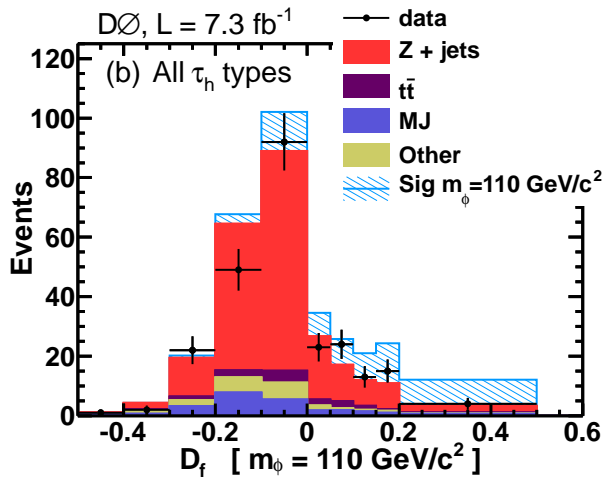}
\caption{ The likeihood discriminant used in the $bh \rightarrow b\tau\tau$ analysis, for a 110 GeV Higgs mass. The stacked histograms show the contribution from the different backgrounds and the hatched histogram show the predicted Higgs signal.} \label{figure_three}
\end{figure}

\subsubsection{ $bh \rightarrow bb\bar{b}$}

For the $bh \rightarrow bb\bar{b}$ analysis a trigger specifically designed to select events with at least 3 jets is used. This trigger is determined to be around 60\% efficient for a 150 GeV Higgs signal when measured with respect to events with 3 or 4 reconstructed jets. 
Events with 3 or 4 jets are selected, 3 of which must satisfy $p_T > 20$ GeV, $|\eta| < 2.4$ criteria and must pass  tight NN $b$ tagging cuts. The efficiency of the $b$ tagging NN is around 50\%, with a fake rate for light jets of around 0.5-1.5\%. An additional cut on the $p_T$ of the two leading jets of greater than 20 GeV is also applied. 
The background is dominated by heavy flavor jets with a smaller contribution coming from light flavor jets. This multijet background is predicted using a  data driven method.  
To enhance the selection of the signal over the background, a likelihood discriminant is used based on a set of kinematic variables for which the background is well modeled.  Two likelihood discriminants are trained, one for the low mass region $90 \leq M_A < 140$ GeV and one for the high Higgs mass region $ 140 \leq M_A < 260$ GeV. The jet pair with the highest likelihood is selected and used to construct an invariant mass which is used as the input to the statistical analysis.  This distribution can be seen in Figure 4. 

\begin{figure}[ht]
\centering
\includegraphics[width=80mm]{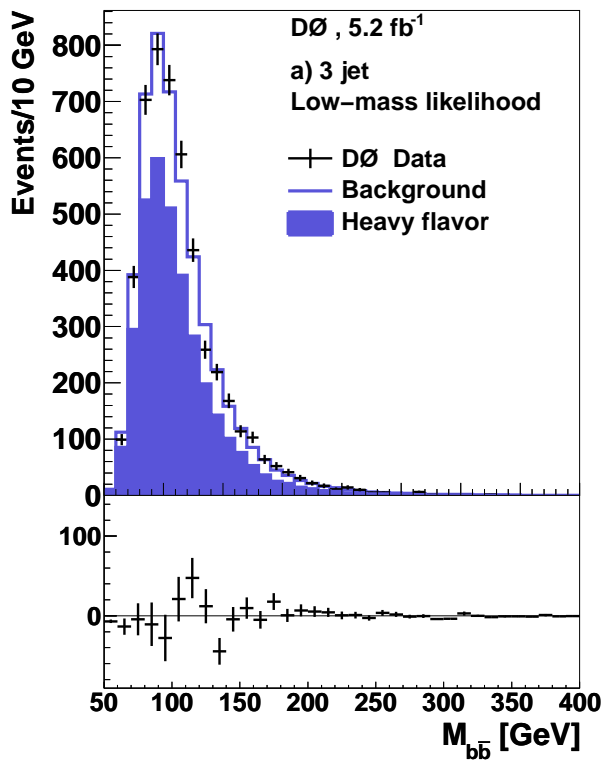}
\caption{ The invariant mass of the highest likelihood pair of $b$ quarks. The line shows the predicted background, with the heavy flavor contribution in blue. The crosses show the data with the difference between the data and the expected background shown below. } \label{figure_four}
\end{figure}

\subsection{Systematic Uncertainties}
Systematic uncertainties are either included as constant systematics on the yield of signal or background events or as a shape systematics, the value of which changes across the different bins of the distribution. 
The dominant sources of systematic uncertainties for $bh \rightarrow b\tau\tau$ analysis are as follows. For the $Z$+jets samples the uncertainness are estimated from $Z \rightarrow \mu\mu$ events,  these include the $Z$+jets normilazation uncertainties, the  inclusive trigger efficiency (applied to all background samples) and a shape uncertainty from the modeling of the Z boson kinematics.
   The other background samples are dominated by the uncertainties on, the hadronic tau reconstruction efficiency,  the luminosity determination, the muon reconstruction efficiency, the single muon trigger efficiency, the $t\bar{t}$ and diboson production cross sections. The determination on the multijet background has a large uncertainty of 10-40\%. Additional dominant shape uncertainties come from the jet energy scale determination and the $b$-tagging efficiency modeling. 
Apart from the uncertainties on the multijet production and the hadronic tau reconstruction efficiency all these uncertainties are taken to be 100\% correlated across the tau types. 

In the case of $bh \rightarrow bb\bar{b}$ only shape systematics are considered. The dominant background arises from the measurement of the rate at which light partons fake  heavy flavor jets and the $b$ tagging efficiency. In contrast, for the signal the dominant sources of uncertainness are the jet energy scale and the $b$ tagging efficiency. 
Between the two analyses only the $b$ tagging efficiency and the jet modeling systematics are correlated and are taken to be 100\% correlated. 

\subsection{Combined Results}

Limits are set using the modified frequentist or CL$_S$ technique ~\cite{cls}, with more details on this available in ~\cite{d0comb}. This is done both within the the framework of the MSSM at the benchmark points as described in Section 2.4 and in a less model independent way on the cross section $\times$ branching ratio. This is not normally possible when combining channels with different production modes or final states, as the relative signal yields in each channel will generally depend on the specific model being studied. But as the two channels share a production mode, and the Higgs width is narrower than the experiment resolution for the majority of the interesting phase space, model independent limits can be derived. The only additional assumption that must be made is that the Higgs boson only decays to tau leptons and $b$ quarks. This model independent limit can be seen in Figure 5, for 3 choices of the tau branching ratio, BR = 6\%, 10\%, 14\%. For masses up to around 180 GeV, the $bh \rightarrow b\tau\tau$ channel tends to dominate, beyond this value the $bh \rightarrow bb\bar{b} $ becomes increasingly important as can be seem  from the decrease in the dependence of the limits on the tau branching fraction at higher values.

\begin{figure}[ht]
\centering
\includegraphics[width=80mm]{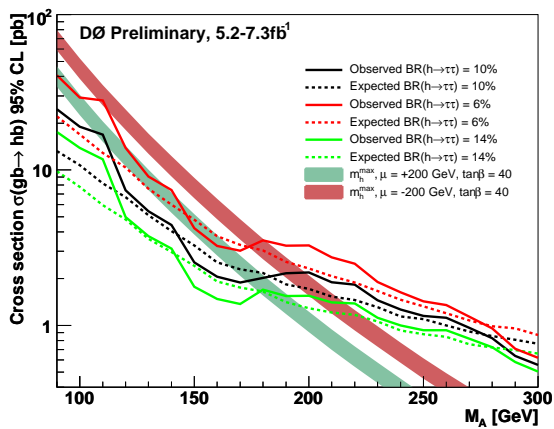}
\caption{ Combined 95\% confidence rates on the cross section. The observed (solid) and expected (dashed) are shown for three assumptions of the tau branching ratio. The shaded bands show the theoretical uncertainty an error of 15\% for the cross section for the $m_h^{max}$ scenario for two different assumptions of $\mu$ and tan$\beta$ = 40.  } \label{figure_five}
\end{figure}

In Figure 6 the limits are set in the $m_h^{max}$ and no mixing scenarios as described in Section 2.4, for a positive value of tan$\beta$. The experimental uncertainties an additional 15\% uncertainly on the production cross section is included. This has contributions from the scale variations and the PDF uncertainties. 
The SM cross section of the $gb\rightarrow hb$ process as taken from MCFM where FEYNHIGGS is used to calculate the branching fraction so that the SM cross section can be corrected to the appropriate $\sigma$ $\times$ BR. For large values of tan$\beta$ the width of the Higgs boson is comparable to the experiment resolution, this effect is taken into account using the method described in ~\cite{widthfix}.
These limits are the best limits so far produced at the Tevatron. 

\begin{figure*}[ht]
\centering
\includegraphics[width=80mm]{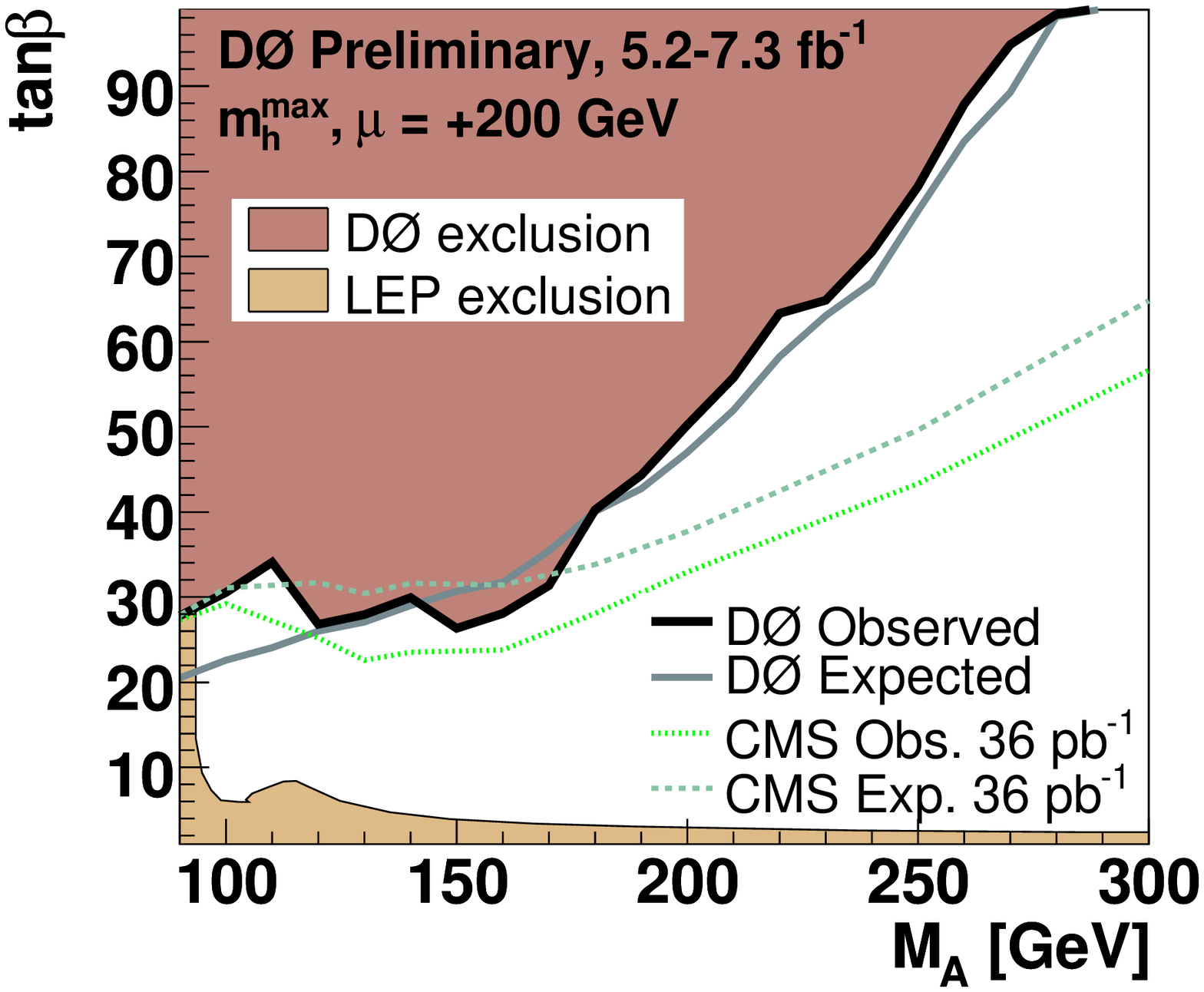}
\includegraphics[width=80mm]{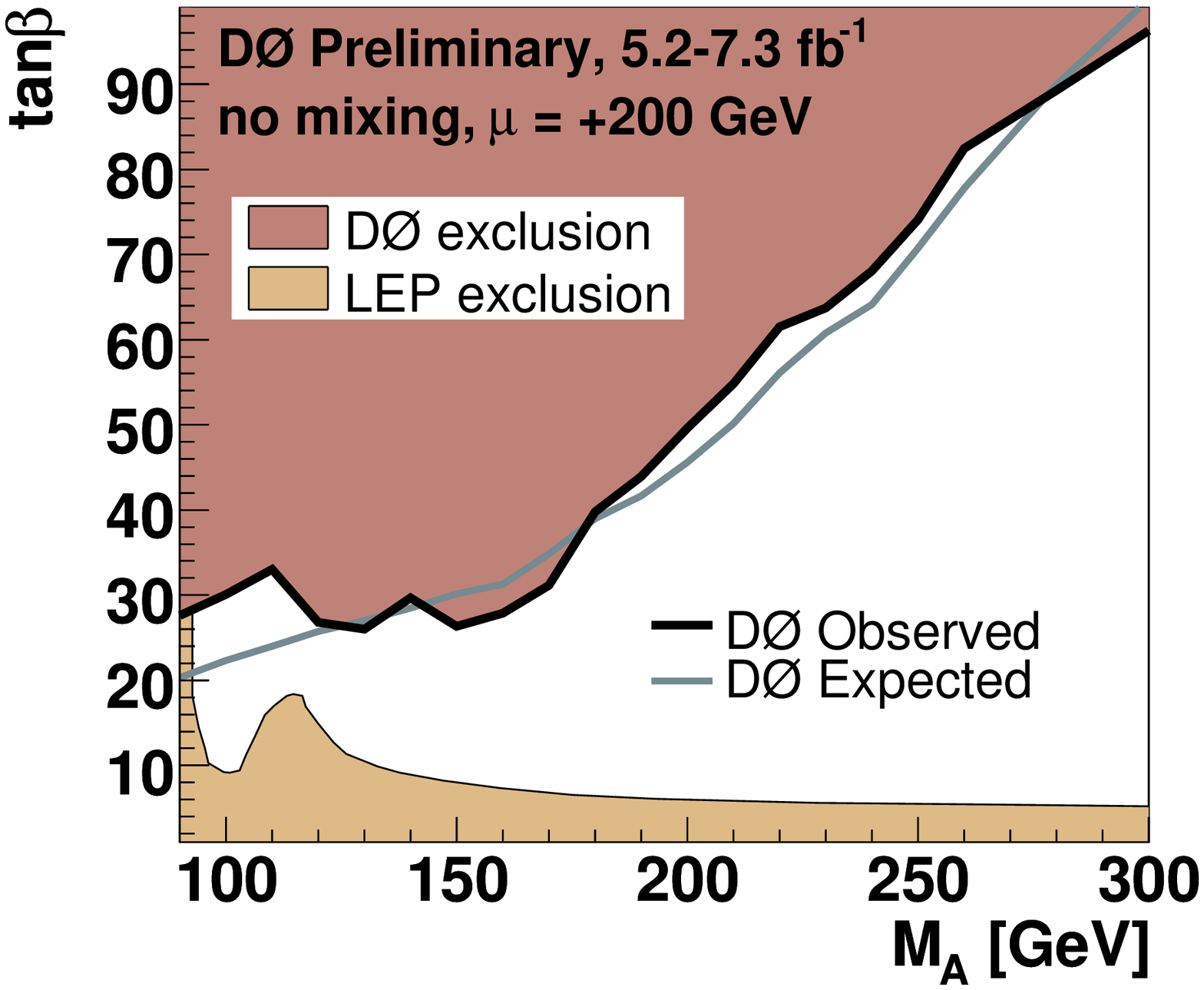}
\caption{ 95\% confidence limits in the $M_A$-tan$\beta$ plane, for the two benchmark scenarios and a positive $\mu$ value. } \label{figure_six}
\end{figure*}


\begin{acknowledgments}
The author would like to thank the organizers of DPF as well as the collaborators  at D0 and CDF who contributed to these analyses. 

\end{acknowledgments}

\bigskip 

\end{document}